\documentclass[useAMS,usenatbib]{mn2e}

\usepackage{times}

\usepackage{graphicx}
\usepackage{amsmath}
\usepackage{amssymb}
\usepackage{natbib}

\newcommand{\apj}{Ap.J.}
\newcommand{\aj}{A.J.}
\newcommand{\mnras}{MNRAS}

\newcommand{\pasp}{PASP}

\newcommand{\nat}{Nature}
\newcommand{\hth}{PaperI}
\newcommand{\thh}{PaperII}

\newcommand{\add}{}

\def\trh0{t_{rh}(0)}

\def\apgt{\ {\raise-.5ex\hbox{$\buildrel>\over\sim$}}\ }
\def\aplt{\ {\raise-.5ex\hbox{$\buildrel<\over\sim$}}\ }

\title[Star Cluster Evolution with Primordial Binaries III]
{Star Clusters with Primordial Binaries:\break
III. Dynamical Interaction between Binaries and an Intermediate Mass Black Hole}
\author[M Trenti, E. ardi, S. Mineshige and P. Hut]
{Michele Trenti$^{1,2}$\thanks{E-mail addresses:
trenti@stsci.edu (MT); eliani@yukawa.kyoto-u.ac.jp (EA); minesige@yukawa.kyoto-u.ac.jp (SH); piet@ias.edu (PH)},
Eliani Ardi$^{1}$\footnotemark[1], Shin Mineshige$^{1}$\footnotemark[1] and Piet Hut$^{3}$\footnotemark[1]\\
$^{1}$Yukawa Institute for Theoretical Physics, Kyoto University,
606-8502 Kyoto, Japan\\
$^{2}$Space Telescope Science Institute, 3700 San Martin Drive, Baltimore, MD, 21218 \\
$^{3}$Institute for Advanced Study, Princeton, NJ 08540, USA}
\begin{document}

\date{Accepted ; Received ; in original form }

\pagerange{\pageref{firstpage}--\pageref{lastpage}} \pubyear{2006}

\maketitle

\label{firstpage}

\begin{abstract}
We present the first study of the dynamical evolution of an isolated
star cluster that combines a significant population of primordial
binaries with the presence of a central black hole. {We use equal-mass
direct N-body simulations, with $N$ ranging from $4096$ to $16384$ and
a primordial binary ratio of $0-10\%$; the black hole mass is about
one percent of the total mass of the cluster}. The evolution of the
binary population is strongly influenced by the presence of the black
hole, which gives the cluster a large core with a central density
cusp. Starting from a variety of initial conditions (Plummer and King
models), we first encounter a phase, that last approximately $10$
half-mass relaxation times, in which binaries are disrupted faster
compared to analogous simulations without a black hole.  Subsequently,
however, binary disruption slows down significantly, due to the large
core size. The dynamical interplay between the primordial binaries and
the black hole thus introduces new features with respect to the
scenarios investigated so far, where the influence of the black hole
and of the binaries have been considered separately. {\add A large
core to half mass radius ratio appears to be a promising indirect
evidence for the presence of a intermediate-mass black hole in old
globular clusters.}

\end{abstract}
%%%%%%%%%%%%%%%%%%%%%%%%%%%%%%%%%%%%%%%%%%%%%%%%%%%%%%%%%%

%%%%%%%%%%%%%%%%%%%%%%%%%%%%%%%%%%%%%%%%%%%%%%%%%%%%%%%%%%
\section{Introduction}
Over the last few years some tantalizing, but yet far from conclusive,
evidence has been accumulating in support of the idea that some star
clusters could harbor a central black hole (BH) with a mass of the
order of $10^3 M_{\sun}$ or more.  Detection of such an intermediate
mass black hole (IMBH) has been claimed for $M15$ and $G1$
\citep{ger03,geb02,geb05}. {\add However, alternative dynamical models
without a central BH have been proposed for these clusters
\citep{bau03a,bau03b}.} Interestingly, the visual appearance of
globulars containing an IMBH is not that of a so-called core-collapsed
cluster, but rather that of a cluster with a still sizable core
\citep{bau04c}.

IMBHs present a high theoretical and observational interest as these
could be potential ultra-luminous X-ray sources and even emit
gravitational waves, detectable by the next generation of
gravitational wave detectors, as a result of close interactions with
stars. However, despite this interest and the fact that theoretical
studies of BHs in stellar systems started more than 30 years ago
\citep[e.g., see][]{pee72,bah76}, detailed direct N-body simulations
to study the dynamics of an idealized model with single stars and a
central BH have been performed only recently (\citealt{bau04a,bau04b};
see also the studies on the formation of IMBHs by runaway mergers of
massive stars by \citealt{por04}).

One ingredient that complicates N-body simulations of globular
clusters is the presence of primordial binaries.  Often, these are
neglected in large simulations despite the increasing observational
evidence that many stars in a globular cluster have a companion
\citep{hut92,alb01,bel02,zha05}.  This frequent neglect is due to the
dramatic increase in computational resources required in a simulation
where the local dynamical timescale may be many orders of magnitude
smaller than the global relaxation timescale (hard binaries have an
orbital period of a few hours, while the half-mass relaxation time can
be up to a few billion years). The study of the dynamics in the
presence of primordial binaries has been mainly limited to
Fokker-Planck or Monte Carlo approaches \citep{gao91,gie00,fre03} and
to direct simulations with rather modest particle numbers, from $N
\approx 10^3$ \citep{mcm90,mcm94,heg92} to recent higher resolution
simulations, with $N$ up to $16384$ (\citealt{heg05,tre06}, hereafter
PaperI and PaperII respectively). {The results of these direct
simulations have to be extrapolated to a realistic number of particles
(e.g., by means of the \citet{ves94} model) before being applied to
the interpretation of the evolution of typical globular clusters, that
have $10^5-10^6$ stars.} Some realistic simulations, including
primordial binaries are available \citep{por04b}, but these are
limited only to the first stage of the life of young dense
clusters. In the case of open clusters, M67 has been modeled in a
36,000-body simulation running for several Gigayears \citep{hur05}.

The presence of either an IMBH or a significant population of primordial
binaries leads to an early release of abundant energy, inhibiting the
development of a deep core collapse and hence of the onset of
gravothermal oscillations.  An IMBH can generate energy by swallowing or
tightly binding stars deep in its potential well [note also that energy
can be generated through encounters of stars in the density cusp that is
formed around the BH], while primordial binaries can generate energy by
rapidly increasing binary binding energy through three and four body
encounters.  Energy thus generated in the core of the system fuels the
expansion of the half-mass radius, leading to a self-similar expansion
of the entire system \citep{hen65}.

In this work we present the first direct simulations ({with a
number of equal-mass particles up to $N=16384$}) of the evolution of
star clusters with both a significant fraction of primordial
binaries (10~\%) as well as an IMBH with a mass of $0.8-3~\%$ of
the total mass of the system. We address the following questions.
Under the combined effect of the BH and of the binaries, what is the
equilibrium size of the cluster core?  How is the binary population
affected by the presence of the central IMBH? Which physical processes
dominate in the core? In principle two competing effects are possible:
the BH may either enhance the disruption rate of binaries both
indirectly, due to the creation of a density cusp, and directly, by
tidal stripping \citep{pfa05}, or it may reduce the probability of
interactions between binaries and singles, by producing a low stellar
density in a relatively large core \citep{bau04a}. Which process is
dominant is of fundamental importance. These questions are addressed
in the next sections.

\section{Numerical simulations: setup}\label{sec:ns}

Like in \hth~and \thh, the simulations (see Table~\ref{tab:runs}) have
been performed using the NBODY-6 code \citep{aar03}. For this project
NBODY6 has been modified with the kind help of Dr. Aarseth to ensure a
more efficient and accurate treatment of the dynamics around the
BH by fine tuning the parameters controlling chain regularization.

\subsection{Units}\label{sec:units}

All our results are presented using standard units \citep{hm86} in which
$$
G = M = - 4 E_{\rm T} = 1
$$ where $G$ is the gravitational constant, $M$ is the total mass, and
$E_{\rm T}$ is the total energy of the system of bound objects. In
other words, $E_{\rm T}$ does not include the internal binding energy
of the binaries, only the kinetic energy of their center-of-mass
motion and the potential energy contribution where each binary is
considered to be a point mass.  We denote the corresponding unit of
time by 
$$
t_d = GM^{5/2}/(-4E_T)^{3/2} \equiv 1.
$$ 
For the relaxation time, we use the following expression \citep{spi87,gie94}:
$$ 
t_{rh} = \frac{0.138 N  r_h^{3/2}}{\sqrt{GM}\ln{(0.11 N)}},  
$$ 
where $r_h$ is the half-mass radius and $N$ denotes the number of
original objects (binaries + singles).

The core radius of the system is defined as the density averaged
radius:
\begin{equation}
 \label{eq:rc} {r}_c=
\sqrt{\frac{\sum_{n=1,{N}}{r_n^2
\rho_n^2}}{\sum_{n=1,{N}}{\rho_n^2}}}, 
\end{equation}
 where $r_n$ is the
distance of the $n$-th star from the density center, and the density
$\rho_n$ around each particle is computed from the distance to the
fifth nearest neighbor \citep{cas85}. A discussion of other possible
definitions for the core radius is extensively given in \citet{cas85};
see also \thh.

Like in \hth~and \thh, the internal energy of the binaries is measured
in units of $kT$, where $(3/2)kT$ is the initial mean kinetic energy
per particle of the system (the binaries being replaced by their
barycenter).

As we are considering simulations with purely classical gravitational
interactions only, all our results are intrinsically scale-free and do
not depend on the specific choices in physical units for the total
mass and for the scale radius of the system modeled. 

For reference in physical units, a globular cluster (described by a
Plummer model) with $N=3 \cdot 10^5$ stars, a total mass of $M=3 \cdot
10^5~M_{\sun}$ and a half-mass radius of $4~pc$ has a half-mass
relaxation time $t_{rh} \approx 8.5 \cdot 10^8~yr$; in this cluster a
binary, formed by equal mass stars each of mass $1~M_{\sun}$, with
binding energy of $1~kT$ has a semi-major axis of $\approx 10~AU$.

\subsection{Initial conditions}\label{sec:icinit}

The models considered in this paper are isolated with stars of equal
mass $m$, plus a BH, introduced as a massive star, with mass
($m_{BH}$) in the range $0.8-3~\%$ of the total mass of the
system. The initial distribution is either a Plummer model (entries
``PL'' in Tab.~\ref{tab:runs}) or a King model (entries ``K'' in
Tab.~\ref{tab:runs}) with concentration index $W_0=3,5,7,11$. Our standard
models have a primordial binary ratio of $0-10~\%$.

%%%%%%%%%%%%%%%%%%%%%%%%%%%
\begin{table}
\begin{center}
\caption{Summary of N-body runs\label{tab:runs}}
\begin{tabular}{lcccc}
\hline\hline
ID & IC type & $N$ & $m_{BH}/M$ & $f$  \\
\hline
Pa & Plummer & 8192 & 0.014 & 0.1 \\
%Pa1 & Plummer & 8192 & 0.014 & 0.1 \\
Pb & Plummer & 8192 & 0.014 & 0   \\
Pc & Plummer & 8192 & 0.025 & 0.1  \\
Pd & Plummer & 8192 & 0     & 0.1  \\
Pe & Plummer & 8192 & 0     & 0   \\
Pf & Plummer & 16384 & 0 & 0.1      \\
Pg & Plummer & 16384 & 0.008 & 0.1     \\
K5a1 & King $W_0=5$ &  8192 & 0.014 & 0.1  \\
K5a2 & King $W_0=5$ &8192 & 0.014 & 0.1    \\
K5b1 & King $W_0=5$ &8192 & 0.014 & 0  \\
K5b2 & King $W_0=5$ &8192 & 0.014 & 0    \\
K7a1 & King $W_0=7$ &8192 & 0.014 & 0.1  \\
K7a2 & King $W_0=7$ &8192 & 0.014 & 0.1    \\
K7b1 & King $W_0=7$ &8192 & 0.014 & 0  \\
K7b2 & King $W_0=7$ &8192 & 0.014 & 0    \\
K3a  & King $W_0=3$ &4096 & 0.03  & 0 \\
K7a  & King $W_0=7$ &4096 & 0.03  & 0\\
K11a & King $W_0=11$ &4096 & 0.03 & 0 \\

\hline
\end{tabular}
%% Any table notes must follow the \end{tabular} command.
\newline {Initial conditions for our set of runs. Columns (left
to right): identification mark for the simulation, type of initial
density profile, number of particles, black-hole mass, fraction of primordial binaries.}
\end{center}
\end{table}
%%%%%%%%%%%%%%%%%%%%%%%%%%%

We define the primordial binary fraction as:
%%%%%
\begin{equation}\label{eq:f}
f = n_b/(n_s+n_b)
\end{equation}
%%%%
with $n_s$ and $n_b$ being the number of singles and binaries
respectively. 

We have considered runs with $N$ in the range $4096-16384$. Note that
$N$ denotes the number of original objects, i.e.  $N=n_s+n_b$, while
the total number of stars is $N_{tot}=n_s +2n_b$: when we discuss a
run with $N=8192$ and 10\% primordial binaries we are dealing with
$9011$ stars.

The initial binding energy distribution for the binaries is in the
range $5$ to $680~kT$, flat in logarithmic scale, similar to the
initial conditions considered in the studies focused on the evolution
of star clusters with primordial binaries, but without an IMBH
(\citealt{gao91,fre03}; \hth; \thh)\footnote{Note that with our definition
of $kT$ we are considering the same range of \citet{gao91}, that
reports the binary binding energy in units of $kT_C$ ($[3kT_c;400kT_c]$),
with a temperature $T_c$ defined in terms of the \emph{central}
velocity dispersion (see Sec.~3.1 in \hth)}. This form for the
binding energy distribution is suggested by the observed properties of
binaries in star clusters \citep{hut92} and the upper and lower limits
are physically motivated: very soft binaries, i.e. binaries with
binding energies below a few $kT$, would be quickly destroyed (i.e. on
a timescale below one relaxation time) due to three and four body
encounters in the core of the the system; binaries harder than a few
hundreds $kT$ are, on the other hand, dynamically inert.

To initialize the simulation in a situation of approximate dynamical
equilibrium in the presence of the BH we first generate our chosen
initial configuration (Plummer or King model) made of $N_{tot}$ single
stars only and we add the IMBH at rest at the center of the system.
We then scale the velocities of the particles to reach virial
equilibrium and let the system evolve for $5$ half-mass crossing times
$t_d$. In this first phase the system settles down in dynamical
equilibrium by experiencing mild virial oscillations (with amplitude
below 6\% of the equilibrium virial ratio) that are damped out in
$2-3~t_d$. At this stage we select randomly $n_b$ single stars that
are eliminated from the simulation. Other $n_b$ single stars are
randomly chosen to become binaries: for these a companion star is
added with a semi-major axis ($a$) extracted from the assumed flat
distribution in binary binding energy (that is translated in a
distribution in $a$ proportional to $1/a$). The initial eccentricity
for the binaries is chosen from a thermal distribution. At this stage
the initialization is complete with the system in approximate
dynamical equilibrium. % and we are ready to start the simulation.

{\add Our initial configurations have a limited choice of initial
density profiles (Plummer model and King profiles with $W_0 =
3,5,7,11$). This does however not influence or bias significantly the
long-term, collisional evolution of the system. In fact, as we have
shown in \thh (see also Sec.~\ref{sec:king}), the memory of initial
conditions is erased within a few half mass relaxation times. Runs
starting from concentrated King models initially presents a core
expansion, while runs starting with shallower cores have an initial
contraction, with both processes leading to a common dynamical
configuration that can be interpreted in terms of balance between
energy production in the core and energy dissipation due to half mass
radius expansion (see also \citealt{ves94}). This argument also
applies to justify our choice in the initialization procedure to
distribute, like in \thh, primordial binaries so as to exactly trace
the distribution of single particles, without exploring the
possibility of a initial mass segregation.}

The evolution of the system is then followed up to $t \approx
25~t_{rh}(0)$ that corresponds to about $20$ Gyga year in physical
units for a globular cluster with $N=3 \cdot 10^5$, $M=3 \cdot
10^{5}~M_{\sun}$ and $r_h=4~pc$ at $t=0$. For $t_{rh}(0)$ we mean
$t_{rh}$ computed at $t=0$. For our initial conditions with $N=8192$,
$t_{rh}(0) \approx 112 t_d$.

At $N=8192$ the CPU time required for this typical run is of the order
of 10 days on a state of the art pc linux workstation. As here we are
interested in the \emph{collisional} evolution of the system our
results are expressed in terms of the relaxation time.  This means
that the computational complexity for direct N-body simulations, if
lasting for a constant number of relaxation times, scales as
$N^3/ln(0.11N)$: the factor $N^2$ comes from the direct gravitational
interaction, while the additional $N/log(0.11N)$ is due to the
increase of the two body relaxation time with respect to the dynamical
time \citep[][Chapter 3]{heg03}.

\subsection{BH properties and $N$ scaling}

The mass of the black hole in our runs is {\add in the range $[0.8 \%;3 \%]$
of the total mass of the system, a value that is slightly higher
than the expected mass for an IMBH following the $m_{BH}-\sigma$
relation \citep{fm00,geb00,tre02} applied to a typical globular
cluster. E.g. using \citet{tre02} with a velocity dispersion of
$10km/s$ we obtain a $m_{BH} \approx 10^3 M_{\sun}$, which is about
$0.3\%$ to $1\%$ of the total mass of a typical cluster. Our choice
for the ratio $m_{BH}/M$} is motivated by the limits that the current
combination of hardware and software set on the number of particles
that can be employed in direct simulations of gravitational systems
with the combined presence of a black hole and primordial binaries.

The number of particles that we can include in one simulation is
between one and two orders of magnitude less than the typical number
of stars in a globular cluster, so we are forced to introduce a
scaling for the mass of the black hole. In principle two possibilities
are available: we can keep constant either the ratio of the black hole
mass to the stellar mass or the black hole mass to the total mass of
the cluster. The first choice has the advantage of keeping unchanged
the local interaction of stars with the black hole, while the second
allows for a better comparison with observed star clusters. With
respect to the properties of the interaction between the black hole
and the primordial binary population it is not clear whether there is
a preferred scaling. In fact, a black hole mass which is too big with
respect to the total mass of the cluster may bias the results of the
simulation, since the black hole may well be more massive than the
core of the system, which usually contains a few percent of the total
mass (see \citealt{bau04a}). But similarly it can be argued that also
a low ratio of the black hole to stellar mass influences the results
of the simulation, as it decreases the efficiency of tidal disruption
of binaries around the black hole cusp, which is proportional to
$m_{BH}/m$. In addition a black hole with a ratio $m_{BH}/m \approx
50$, i.e. a $1~\%$ mass black hole in a system with $5~10^3$
particles, has a small, but non negligible, probability of being
ejected from the core as a result of a strong interaction with a
binary. E.g. this happened during some test runs with $m_{BH}=0.01$
and $N=4096$ on a timescale of about $t=20~t_{rh}(0)$, so that we were
forced to stop these simulations at the moment of the ejection.

Finally here we consider the black hole as a massive particle, and no
mass accretion due to tidal disruption of stars is taken into
account. However, this is not expected to bias the results
significantly: in fact (1) the mass accretion, whose rate is
proportional to the ratio $(m_{BH}/m)^{61/27}$, is not very important
in runs with $N \lesssim 10^4$ (see \citealt{bau04a} Table 1); in
addition (2) stars passing close to the black hole may be captured, so
that they become inert with respect to the other stars (effectively
behaving like as if they had been accreted onto the black hole) or
ejected from the system at extremely high speed (above 100 times the
mean velocity dispersion), so that also in this case the star is
removed from the system (with the difference that its mass is not
added to that of the BH).

\section{Global evolution}\label{sec:sim}

After the initialization described in Sec.~\ref{sec:icinit}, our
simulations start from a condition of dynamical equilibrium: the
evolution of the system is due to collisional effects and happens on a
$t_{rh}$ timescale. This means in particular that the system is in
virial equilibrium ($2K/U = 1$, where $K$ is the kinetic energy and
$U$ is the potential energy of the system). Only occasionally there
are fluctuations of $2K/U$ at the order of one percent, when a strong
interaction between a hierarchical system (e.g. a binary interacting
with the IMBH) takes place leading to the consequent ejection of a
high velocity particle from the system. The virial ratio returns to
the equilibrium value on a dynamical time-scale.

\begin{figure}
\resizebox{\hsize}{!}{\includegraphics{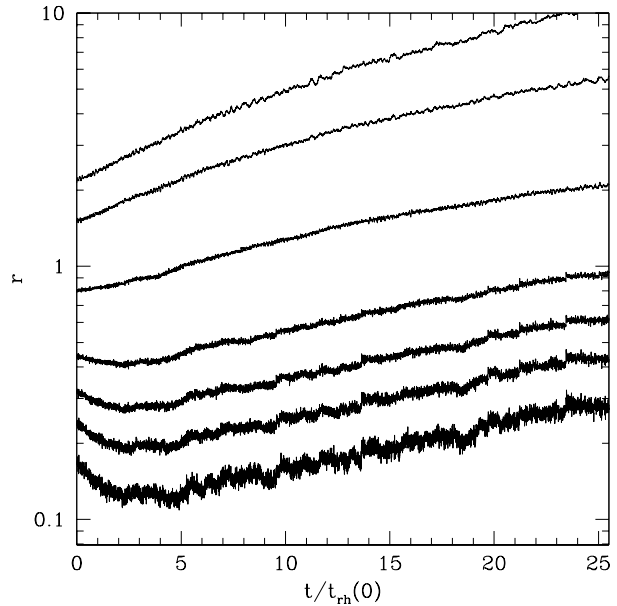}} 
\caption{Lagrangian radii (enclosing from bottom
to top 2,5,10,20,50,80 and 90 \% of the total mass of the stars of the
system) for a run with N=8192 starting from a Plummer profile with
$f=10\%$ and $m_{BH}=0.014$.\label{fig:lrpl}} 
\end{figure}

\begin{figure}
\resizebox{\hsize}{!}{\includegraphics{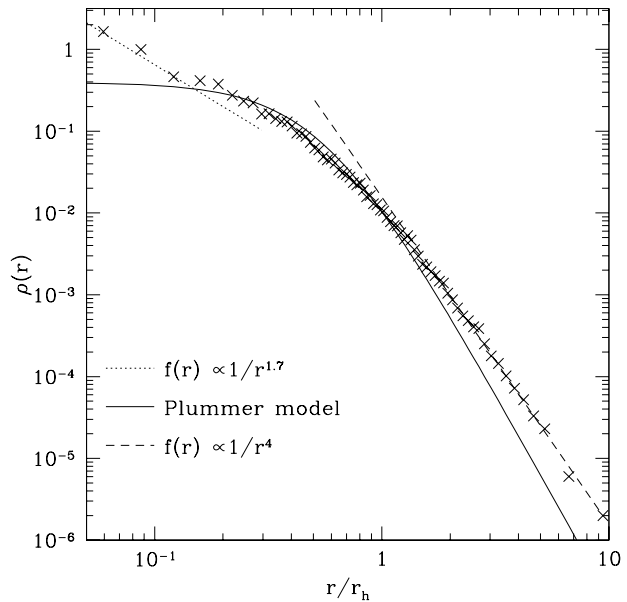}} 
\caption{Density profile (in natural units) vs radius (in units of the
  half mass radius $r_h=1.44$) at time $t=13~t_{rh}(0)$ for a run with
  N=16384 starting from a Plummer profile with $f=10\%$ and
  $m_{BH}=0.014$. {\add The plot shows the stellar mass only. The BH
  contribution to the plot would be that of a delta function located
  at $r=0$. The large core size, basically that of a Plummer model, is
  an indirect consequence of the presence of the central IMBH.}
  \label{fig:rho}}
\end{figure}

The large scale structure evolution of the cluster is dominated by the
heating related to the presence of the BH. Starting from a Plummer
model only the inner regions of the system experience a mild collapse
on a time-scale that, depending on the mass of the BH, is of the order
of a few $t_{rh}$ (depicted in Fig.~\ref{fig:lrpl}). Inside the core
radius $r_c$ a cusp in the density profile is formed within the sphere
of influence $r_i$ (with $r_i \approx 15~r_c m_{BH}/M$, see
\citealt{bau04a}) of the BH, with a profile proportional to $\approx
1/r^{1.7}$ (see fig.~\ref{fig:rho}) and thus similar to the
$1/r^{1.75}$ measured by \citet{bau04a}. For $m_{BH}=0.014M$ the
influence radius is approximately $0.2r_c$.  By definition, the
stellar mass within this radius is comparable to that of the BH, and
thus around one percent of the total mass of the cluster.

{\add The limited number of particles that we employ does not allow us to
significantly characterize the anisotropy profile of the system, that
appears to be consistent with quasi-isotropy, except for the outer parts of
the system, where there is a mild excess of radial orbits. As the
presence of a population of binaries does not influence, to a first
approximation, the evolution of the anisotropy profile, we refer to
the discussion in \citet{bau04a} (see in particular their Fig. 8),
where the authors take advantage of the absence of primordial
binaries and run simulations with $O(10^5)$ particles.}

%%%%%%%%%%%%%%%%%%%%%%
%%%
%%%%%%%%%%%%%%%%%%%%%

\subsection{Plummer Models}

For a Plummer model with $N=8192$ and $f=10\%$, in case of
$m_{BH}=0.014M$, the core radius $r_c$ is reduced in $\approx
4~t_{rh}(0)$ from the initial value of $0.4$ to $0.3$ (in natural units,
see Sec.~\ref{sec:units}). This is to be compared with a value of
$\approx 0.1$ reached without the BH (\hth; see also
Tab.~\ref{tabPL}); when a BH (with the same mass) but no
primordial binaries are present $r_c$ goes down to $\approx 0.28$
. From these values it is apparent that the IMBH dominates the global
evolution of the system, as there is very little difference in
simulations with and without primordial binaries if an IMBH is
present.

%%%%%%%%%%%%%%%%%%%%%%%%%%%%%

\begin{table}
\begin{center}
\caption{Global properties of Plummer runs with $N=8192$\label{tabPL}}
\begin{tabular}{ccccc}
\hline\hline
$m_{BH}/M$ & $f$ & $(r_c)_{cc}/r_{c0}$ & $r_c/r_h$ & $r_{hf}/r_{h0}$ \\ 
\hline
0.014 & 0.1 & 0.75 & 0.29 & 2.46 \\
0.025 & 0.1 & 0.80 & 0.31 & 2.83 \\
0.014 & 0 & 0.70 & 0.27 & 1.92 \\
0 & 0.1 & 0.25 & 0.09 & 2.25 \\
0 & 0   & 0.02 & 0.015 & 1.98  \\
\hline
\end{tabular}
%% Any table notes must follow the \end{tabular} command.
\newline {In the first column we report the BH mass, in the second
 the fraction $f$ of primordial binaries, in the third the core radius
 at the end of the initial core contraction phase $(r_c)_{cc}$ in units
 of the initial core radius $r_{c0}$, the fourth entry is the core to
 half mass radius ratio during the self-similar expansion of the system,
 while the last entry ($r_{hf}$) is the value of the half mass radius at
 $t=24~t_{rh}(0)$ in units of the initial half mass radius.}
\end{center}
\end{table}
%%%%%%%%%%%%%%%%%%%%%%%%%%%%%

After the first mild contraction, all Lagrangian radii (shown in
Fig.~\ref{fig:lrpl}) start to expand steadily and a self-similar
regime sets in, with the half-mass radius growing in proportion to
$\approx t^{2/3}$, in agreement with the theoretical argument given by
\citet{hen65} and as found for runs with single stars only by
\citet{bau04a}. The rate of expansion of the half-mass radius is set
by the amount of energy generated in the core \citep{ves94}. In our
runs with binaries plus IMBH this is marginally bigger (by $\approx
10~\%$ at $24~t_{rh}(0)$) than in runs with binaries only; without a BH
the half-mass expansion starts only after core collapse, which takes
$\approx 10~t_{rh}(0)$ in that case (\hth). Conversely, if a BH
but no binaries are present, the expansion rate of $r_h$ is reduced by
$\approx 20~\%$. A summary of the properties of the Plummer runs is
reported in Table~\ref{tabPL}.

\begin{figure}
\resizebox{\hsize}{!}{\includegraphics{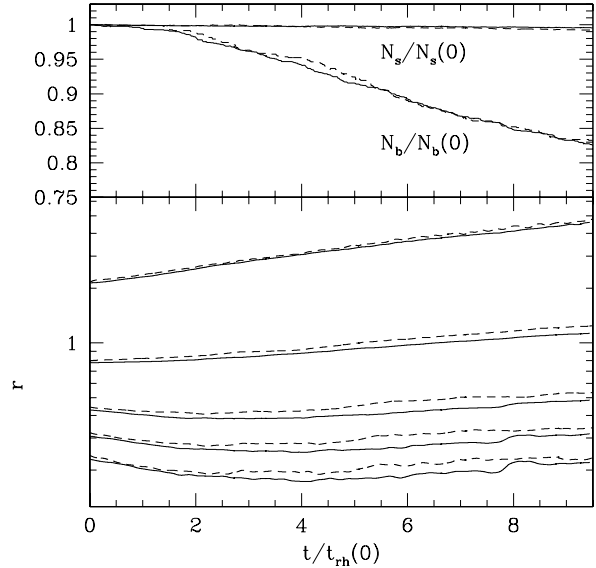}} 
\caption{Comparison between two simulations starting from a Plummer
  profile with $f=10\%$: one has $N=16384$ with $m_{BH} = 0.008$
  (solid line), the second has $N=8192$ with $m_{BH}=0.014$ (dashed
  line). In the upper panel we depict the relative number of binaries
  and singles, in the lower panel selected lagrangian radii (enclosing
  $5,10,20,50$ and $80~\%$ of the total mass in stars). The Lagrangian
  radii have been smoothed with a triangular window filter of width
  $1.5~t_{rh}(0)$. Doubling the number of particles does not introduce
  significant changes in our simulations. There is only a small
  contraction of the Lagrangian radii, when the BH mass is reduced to
  $0.008$.\label{fig:validation}}
\end{figure}

This picture does not appear to depend much on the number of particles
used, except quantitatively for the known logarithmic dependence on
$N$ of the ratio $r_c/r_h$ \citep{ves94}. To validate the results of
our series of $N=8192$ runs we have performed a run with $N=16384$
and an IMBH mass $m_{BH}=0.008$ (thus keeping quasi-constant the
ratio $m/m_{BH}$ and going toward a BH mass closer to the one expected
by the $m_{BH}-\sigma$ relation). The evolution of the system, reported in
Fig.~\ref{fig:validation}, is indeed very similar, both in terms of
binary destruction rate and in terms of the Lagrangian radii
evolution.

%%%%%%%%%%%%%%%
%%%%%%%%%%%%%%%
\begin{figure}
\resizebox{\hsize}{!}{\includegraphics{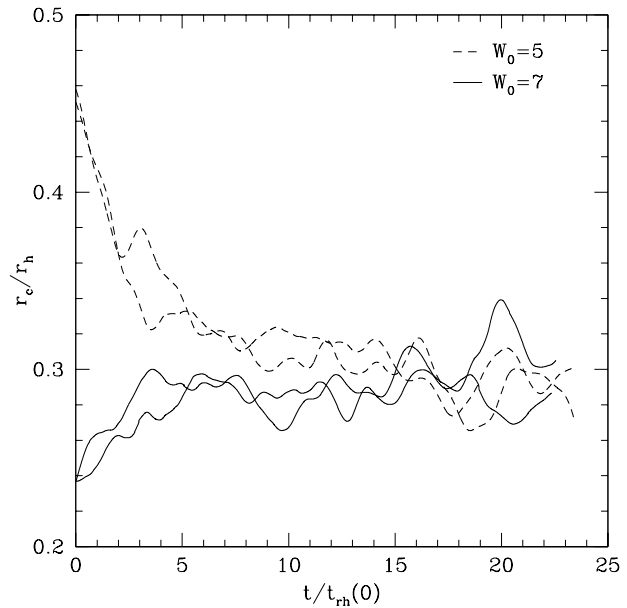}}
\caption{Core to half mass radius ratio for a
  series of simulations with $N=8192$, $m_{BH}=0.014$ and $10\%$
  primordial binaries. Different initial conditions (King profiles
  with $W_0=5,7$) converge toward a common value for
  $r_c/r_h$. The results are presented by applying a triangular
  smoothing window of width $1.5 t_{rh}(0)$. \label{fig:rcrh}}
\end{figure}
%%%%%%%%%%%%%%%%
%\begin{figure}
%\resizebox{\hsize}{!}{\includegraphics{f5.eps}}
%\caption{Like Fig.~\ref{fig:rcrh} but for a
%  series of simulations with $N=4096$, $m_{BH}=0.03$ and no primordial
%  binaries. Also for these conditions the evolution of models starting
%  from very different initial conditions (King profiles with
%  $W_0=3,7,11$) leads to a common state after a few relaxation
%  times.  The results are presented by applying a triangular
%  smoothing window of width $2.5 t_{rh}(0)$. \label{fig:bhcore}}
%\end{figure}
%%%%%%%%%%%%%%%%

\subsection{King Models}\label{sec:king}

At fixed number of particles, IMBH mass and number of primordial
binaries, there is no expectation that the long term evolution of the
system depends much on the details of the initial conditions, as these
are erased on a relaxation time-scale {\add (see also \thh)}. To
verify this we have also started some of our runs (see
Table~\ref{tab:runs}) from an initial King profile, with different
values of the concentration parameter $W_0$. The evolution of the core
radius with respect to the half mass radius is reported in
Fig.~\ref{fig:rcrh}. Both the $W_0=5$ and the $W_0=7$ models evolve
toward $r_c/r_h \approx 0.3$ on a time-scale of $\approx 5~t_{rh}(0)$
(this is the same value reached in the simulations starting from a
Plummer profile). Interestingly the $W_0=7$ model presents an initial
expansion of the core, as the initial central concentration is so high
that the energy released by three and four body encounters cannot be
completely dissipated by expansion of the half mass radius. This
behavior is similar to that observed in runs with primordial binaries
only but with yet higher central concentrations (\citealt{fre03};
\thh). This picture is confirmed also in a different series of runs -
starting from $W_0=3,5,7,11$ with $m_{BH}=0.03$ and no binaries -{\add
that exhibit a very similar behavior.}
 
\section{Properties of the binary population}\label{sec:bin}

{\add The evolution of the number of binaries in our simulations is
driven by two competing processes: the formation of new binaries and
the destruction of existing ones. As we discussed in detail in
\hth~(see in particular Sec.~2 in that paper), the probability of
forming new binaries is greatly suppressed with respect to the
probability of destructing an existing binary. In fact, the leading
formation channel for a new binary from single stars is a three body
encounter, which is proportional to $\rho_s^3$. $\rho_s$ is the number
density of single stars and this density is expressed in physical
units, appropriate for encounters, such that $\rho = 1$ corresponds to
inter-particle distances equal to the ninety-degree turnaround impact
parameter (\citealt{hut89}). Therefore in this units we have $\rho_s
\ll 1$. The leading destruction channels are instead given by
binary-single and binary-binary encounters and therefore proportional
to $\rho_b \rho_s$ and $\rho_b^2$, where $\rho_b$ is the number
density of binaries, in the same physical units considered for
$\rho_s$. As during our simulations the core number density of
binaries is similar to that of singles, these two last channel
dominates over the three single stars encounters. This expectation is
confirmed empirically in our runs, where the number of dynamically
formed binaries that can be found after a few relaxation times in the
simulation (typically about $10^{-3}$ of the number of single stars)
is negligible not only with respect to the number of primordial
binaries but also with respect to the number of disrupted
binaries. Essentially, the dominant dynamical mechanism that changes
the net number of binaries is that of binary disruption. Of course,
the exchange of binary members during a binary-single encounter is
also frequent, but this does not alter the net number of binaries in
the cluster.}

As the presence of the BH dominates the global dynamics of the
cluster, the evolution of the binaries presents some remarkable
differences from the scenario where no central BH is present.  We can
distinguish two phases.

\begin{figure}
\resizebox{\hsize}{!}{\includegraphics{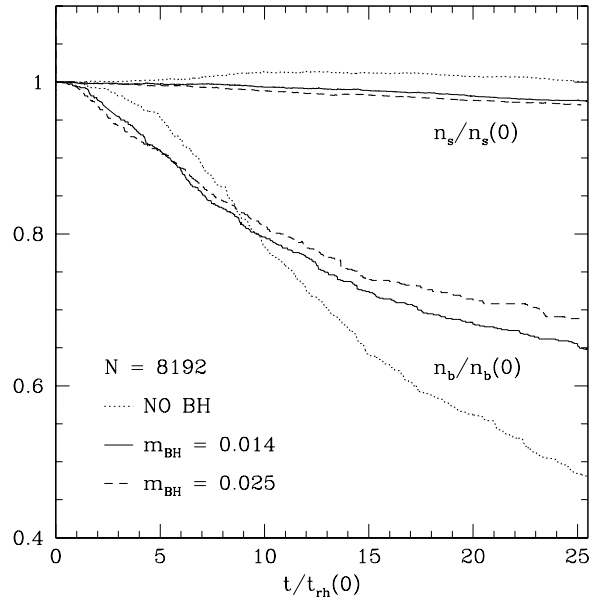}}
\caption{Time-dependence of the number of
single (upper set of lines) and binary stars (lower set of lines)
expressed as a fraction of the initial values. The data refer to
simulations with $N=8192$ and $10\%$ primordial binaries.  There are
three choices for BH mass: no BH
(dotted line), a central BH with $m_{BH}=0.014M$ (solid line),
and $m_{BH}=0.025M$ (dashed line).  The unit of time is the initial
half-mass relaxation time.\label{fig1}}
\end{figure}

The presence of a BH significantly accelerates the rate at which
binaries are disrupted in the first few half-mass relaxation times
(see Fig.~\ref{fig1}).  The reason is that binaries
that pass near the BH can be quickly destroyed. The disruption rate is
approximately constant during the first $5~t_{rh}(0)$, while without a
BH disruption takes a while to get underway. The initial disruption
rate depends on the mass of the BH: a more massive BH starts burning
binaries at a higher rate, as can be seen from
Fig.~\ref{fig1}.

%%%%%%%%%%%%%%%%%%%%%%%%%%%%%%%%%%
%%%%%%%%%%%%%%%%%%%%%%%%%%%%%%%%%%%%%%%%%%%%%%%%%%%

For our Plummer run with $N=8192$, $m_{BH}=0.014 M$ and $f=10\%$ we
observe a binary depletion rate of $\approx 0.15$ binaries per initial
half-mass crossing time ($t_d$), during the first $5~t_{rh}(0)$ (here
$t_{rh}(0)=116.1 t_d$).

\begin{figure}
\resizebox{260pt}{!}{\includegraphics{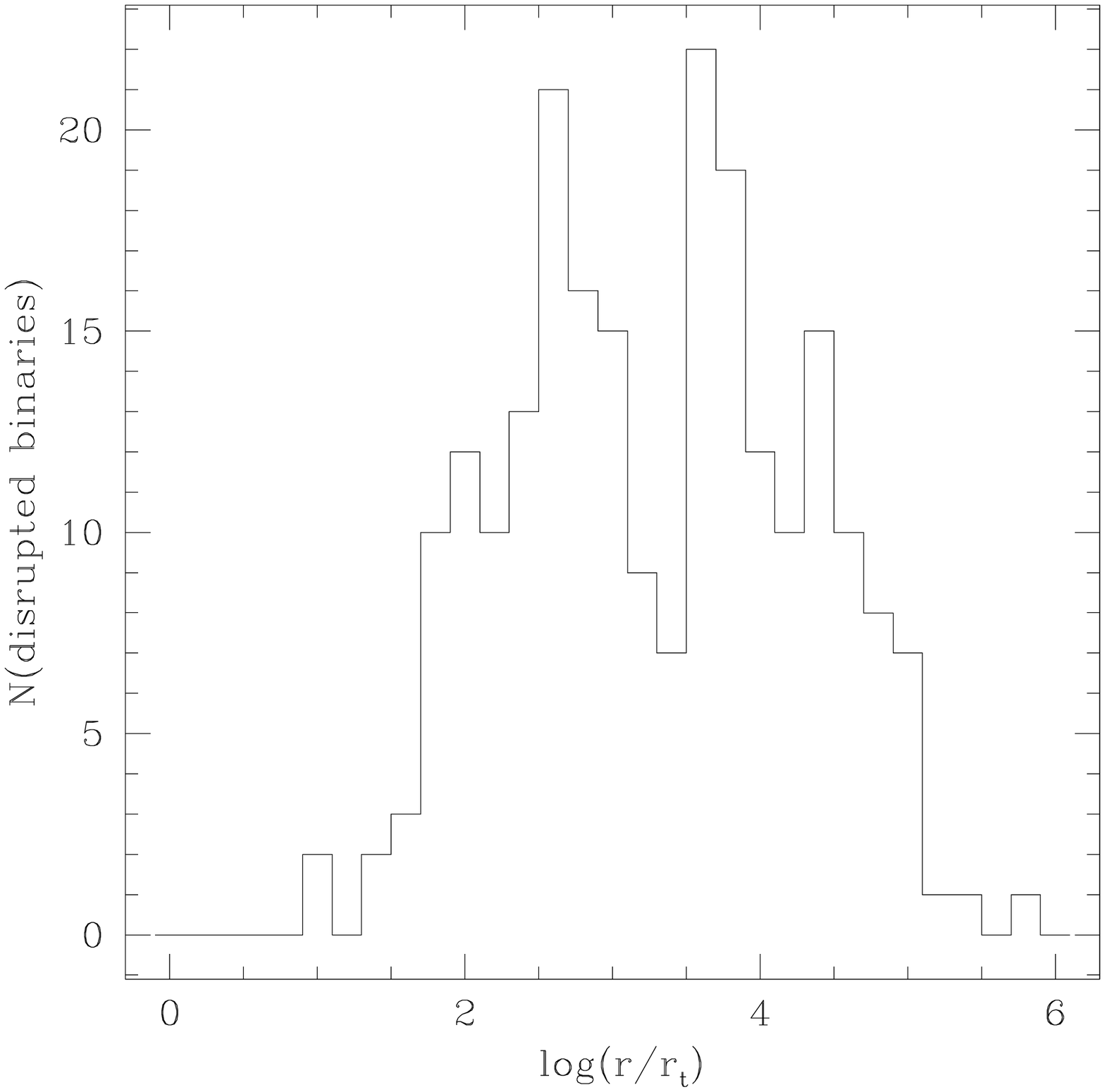}\includegraphics{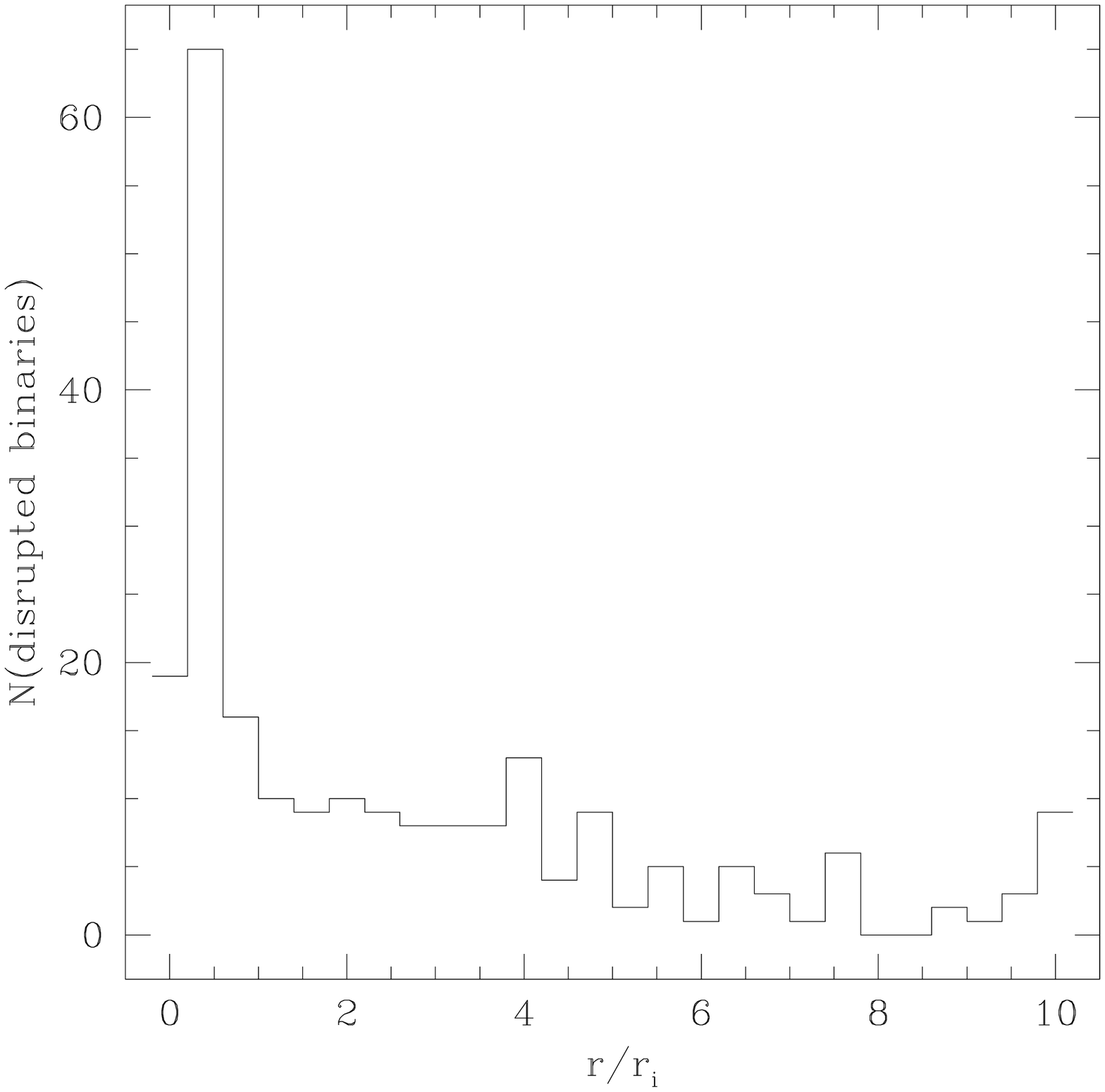}} 
\caption{ Location in the cluster of the
disruption of hard binaries (measured by KS regularization
terminations) during the first $5~t_{rh}(0)$ of a run with $N=8192$ and
$10\%$ primordial binaries starting from a Plummer profile. In the
left panel the radius is given in units of the local tidal radius for
each disrupted binary ($r_t$), while in the right panel the radius is
in units of the influence radius of the IMBH ($r_i$).\label{fig:RIRT}}
\end{figure}

For this time interval we present in Fig.~\ref{fig:RIRT} the
distribution of the radial positions at which binaries are
destroyed. This is obtained by looking in NBODY6 at the position at
which binaries are flagged by the code as too wide to continue meriting
special numerical treatment in terms of KS \citep{ks64} solutions (see
\citealt{aar03} for a detailed description of the algorithm). In
short, a KS solution is terminated in case of an external perturbation
at the level of $5\%$ or higher of the internal gravitational force
between the members of the binary (if the binary is not destroyed by
the perturbation, its special numerical treatment is eventually
restarted when the external force returns below the critical $5\%$
threshold). In the histogram of Fig.~\ref{fig:RIRT} there are 226
events registered, but these include also close encounters with
perturbers that do not change the net numbers of binaries
(e.g. exchanges). The net number of destroyed binaries for the time
interval considered is $80$ (see Fig.~\ref{fig1}). Around half of the
binaries are disrupted within the black hole sphere of influence $r_i$
($100$ out of $226$ KS solutions terminated within
$r_i$). Interestingly only two of them happen to be so close to the BH
tidal radius $r_t$ that the disruption may be considered related to
the tidal stripping force exerted by the BH (see left panel of
Fig.~\ref{fig:RIRT}). Here the tidal radius is defined as the radius
where the gravitational force between the members of the binary equals
the tidal force exerted by the BH (see e.g., \citealt{spi87}), i.e. as
$$
r_t=(m_{BH}/m)^{1/3} a,
$$ where $a$ is the semi-major axis of the binary.  With this respect,
the theoretical model recently proposed by \citet{pfa05} applied to
our $N=8192$, $m_{BH}=0.014$, $f=10\%$ simulation would give (from his
Eqs.~11-12) a tidal disruption rate of $\approx 2 \cdot 10^{-2}$
binaries per initial half-mass crossing time, so that we observe less
tidally stripped binaries than expected ($\approx 10$).

\begin{figure}
\resizebox{260pt}{!}{\includegraphics{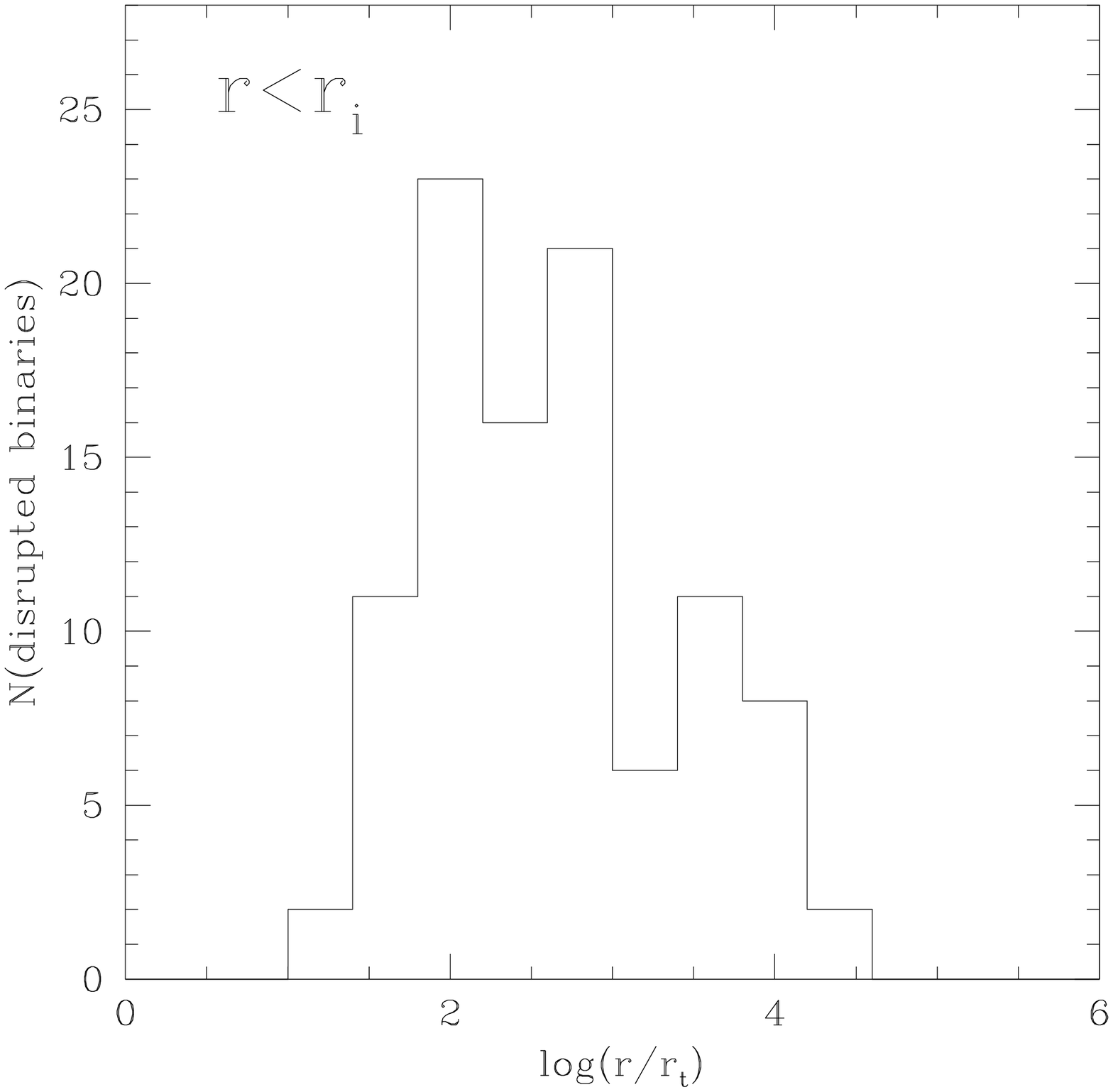}\includegraphics{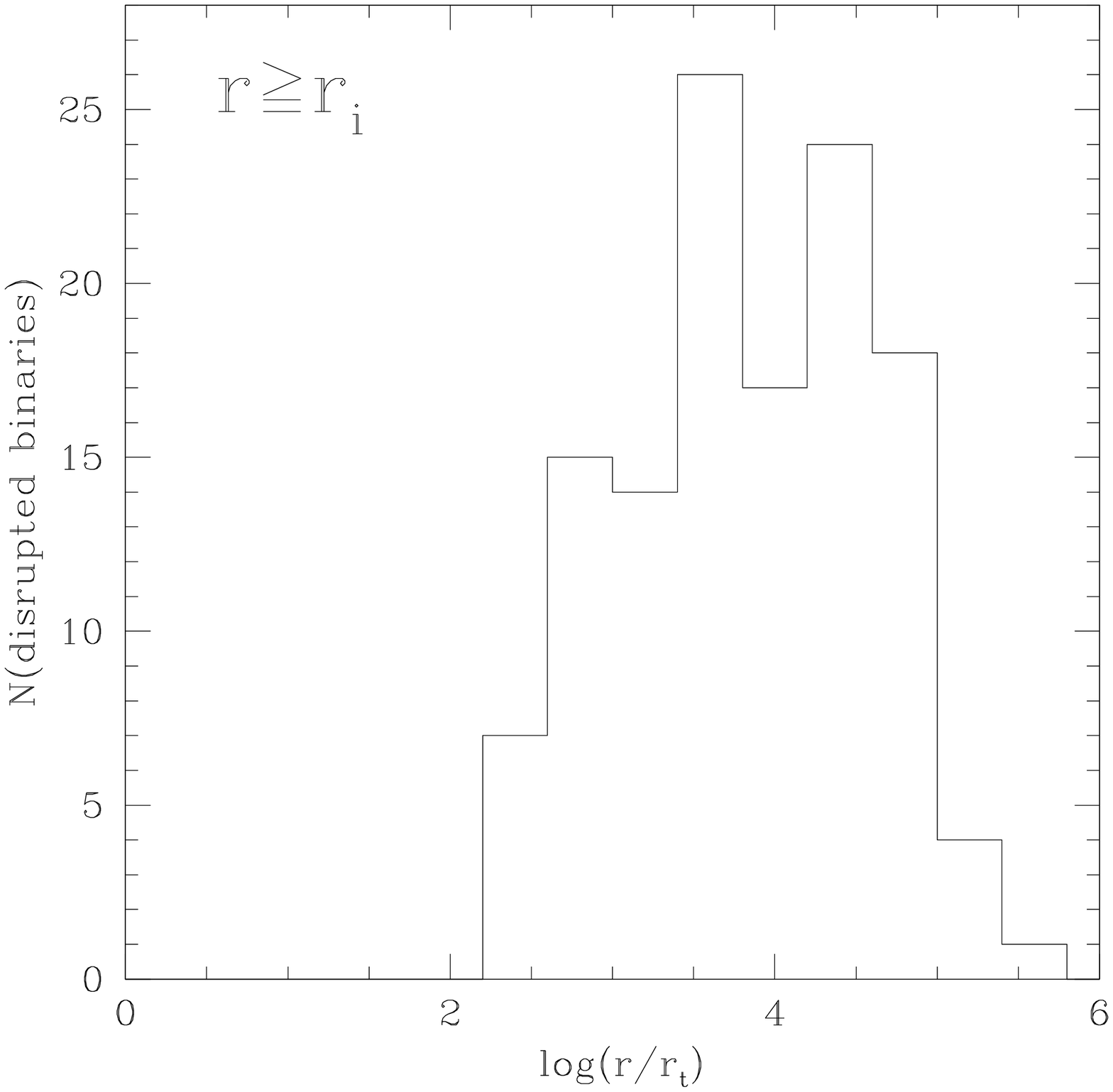}}  \caption{Decomposition of the left panel of
Fig.~\ref{fig:RIRT} into binaries disrupted inside (left) and outside
(right) the influence radius $r_i$ of the BH. The origin of the
bimodal distribution in Fig.~\ref{fig:RIRT} can be undestood in terms
of an enhanced probability for a binary of experiencing a close
encounter when located inside the influence sphere of the
BH.\label{fig:decom}}
\end{figure}

The disruption of binaries by the BH is thus mainly due to an indirect
effect: binaries venturing close to the BH, where the density is
higher, are more likely to interact with a single star or with another
binary before having the chance to being tidally destroyed by the
IMBH. In fact in our simulations we often observe the presence of
hierarchical systems within $r_i$, with the BH playing the role of a
perturber. Fig.~\ref{fig:RIRT} nicely summarizes this picture:
binaries are disrupted at a significantly enhanced rate within $r_i$
(right panel), but the minimum distance from the BH is the order of
$10~r_t$ (left panel). The bi-modal distribution in the left panel of
Fig.~\ref{fig:RIRT} also reflects the enhanced probability of close
encounters and binary disruption within the influence sphere of the
BH, as can be seen from Fig.~\ref{fig:decom}.

After the initial transient phase a self-similar expansion sets in,
where the average core density is much less (approximately by a factor
$10$) than it would be without the presence of a central compact
object.  In this second phase we observe (see Fig.~\ref{fig1}) a
reduced rate of binary disruption (as this rate is proportional to the
square of the density, e.g. see \citealt{ves94}), so that the
difference between a simulation with and without a BH becomes
remarkable. The turning point is around $10~t_{rh}(0)$, when the number
of surviving binaries for a simulation with a central object becomes
greater than in the absence of a BH. Our Fig.~\ref{fig1} has been
given in units of the initial half-mass relaxation time, but the
picture remains qualitatively the same even if we consider a co-moving
time coordinate to take into account the differences between the
half-mass radii of a simulation with and without a BH.

\begin{figure}
\resizebox{\hsize}{!}{\includegraphics{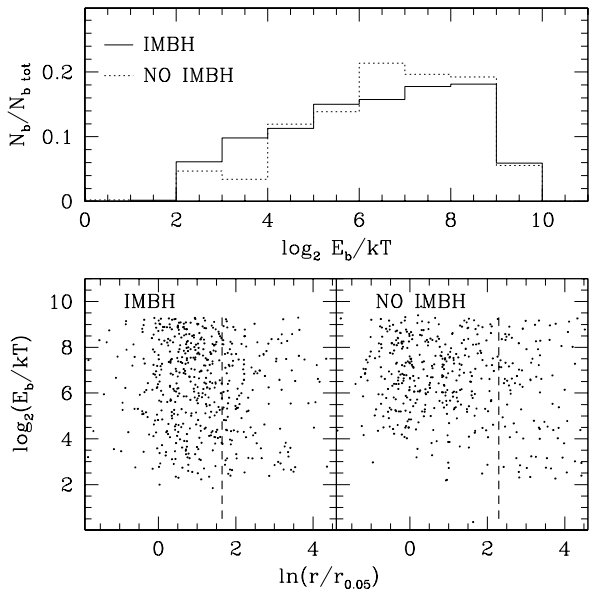}} \caption{Distribution of binding energies for
binaries (upper panel) at time $t=24~t_{rh}(0)$ for a simulation
starting from a Plummer model with $N=8192$, $10~\%$ primordial
binaries and $m_{BH}=0.014M$ (solid line), compared to a similar
simulation without the BH (dotted line). In the lower two panels we
report the energy-radius distribution of binaries for the same
simulations. The radii have been normalized to the Lagrangian radius
containing $5~\%$ of the mass in the simulation. The dashed line gives
the position of the half-mass radius, at $t=24~t_{rh}(0)$. The
relatively large survival at $t=24~t_{rh}(0)$ of binaries within the
half-mass radius with binding energy $E_b<20kT$ is clearly visible in
the bottom panel, and is also reflected in the higher values for the
solid line at the left-hand side of the upper panel.\label{fig2}}
\end{figure}

Interestingly the spatial distribution of binaries is also different
from the case without a BH (see Figs.~\ref{fig2}). After $\approx
20~t_{rh}(0)$ the number of binaries within the $0.05$ Lagrangian
radius is much less for the simulation with a BH. This is probably due
to the disruption of binaries that approach close to the BH. In the
presence of a BH, binaries tend to be more concentrated between the
$0.05$ Lagrangian radius and the half-mass radius, while in the
absence of a BH, binaries can sink deeper into the central region of
the cluster.  As noted by \citet{bau04a} for simulations without
primordial binaries, the mass segregation efficiency is different in
presence of a central BH.

If we compare the evolution of the binding energy distribution of the
binaries in a run with and without a central BH, we can see that the
net effect of the BH at later times is to somewhat enhance the
survival probability of binaries with binding energies of a few $kT$s
($E_b \lesssim 16~kT$), especially for binaries within the half-mass
radius (see Fig.~\ref{fig2}). This feature is especially apparent in a
two-dimensional plot of the distribution of binaries as a function of
binding energy and radial distance from the cluster center, depicted
in the bottom panels of Fig.~\ref{fig2}. The rather strong correlation
between radius and binding energy that is observed \citep[e.g.,
see][\hth]{gie00} in systems without a central compact object is thus
almost absent. The ratio of moderately hard ($E_b/kT \leq 16$) to hard
($64 \leq E_b/kT \leq 128$) binaries is greater by about a factor $2$
at $t \approx 20~t_{rh}(0)$ in our runs with the presence of an IMBH
(see top panel of Fig.~\ref{fig2}). {\add The possible use of this
ratio as a diagnostic to evidence the presence of an IMBH can be
impaired if there is an initial segregation of primordial binaries
depending on their semi-axis. Specifically if softer binaries are
preferentially born in the core of the cluster then some of them could
still survive after a few relaxation times, mimicking the signal due
to an IMBH. However such a cluster would be characterized by a
significantly smaller core radius with respect to a cluster truly
hosting an IMBH.}

\begin{figure}
\resizebox{\hsize}{!}{\includegraphics{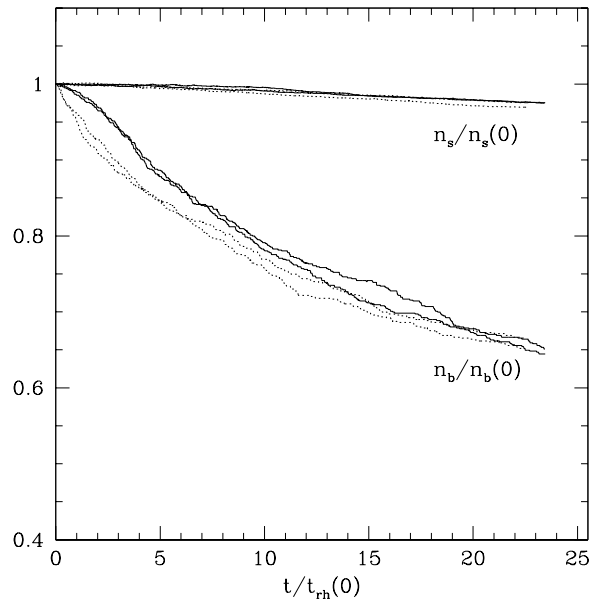}} \caption{Like Fig.~\ref{fig1} but for a series
  of simulations ($N=8192$, $f=10\%$, $m_{BH}=0.014$) starting from
  King models with $W_0=5$ (solid line; two different runs) and
  $W_0=7$ (dotted line; two different runs). After an initial
  transient given by the different initial conditions, the binary
  destruction rate is very similar among all the runs and is
  quantitatively consistent with the one observed in Fig.~\ref{fig1}
  for an initial Plummer density profile. \label{fig:bin_king}}
\end{figure}

This picture, presented for a $N=8192$ Plummer model, is
representative of all our set of simulations. We have shown in
Sec.\ref{sec:sim} that the global evolution of the system is not
sensitive on the details of the initial conditions. Similarly the
evolution of the binary abundance at fixed relaxation times does not
significantly depend on the number of particles $N$ (see
Fig.~\ref{fig:validation}) or on the details of the initial
conditions: in Fig.~\ref{fig:bin_king} we depict the evolution of
$N_{bin}/N_{bin}(0)$ for a number of different simulations and the
pattern is consistently similar. {\add The same trend shown in
Fig.~\ref{fig2} is also obtained starting from different initial
conditions, as we have verified in the runs starting from $W_0=5$ and
$W_0=7$ King models.} All these results show that the evolutionary
differences in runs with and without a central black hole are physical
and not just due to random fluctuations.

%%%%%%%%%%%%%%%%%%%%%%%%%%%%%%%%%%%%%%%%%%%%%%%%%

%%%%%%%%%%%%%%%%%%%%%%%%%%%%%%%%%%%%%%%%%%%%%%%%%%%

%%%%%%%%%%%%%%%%%%%%%%%%%%%%%%%%%%%%%%%%%%%%%%%%%%%%%

%\begin{figure}
%\resizebox{\hsize}{!}{\includegraphics{f11.eps}} \caption{Like Fig.~\ref{fig2} but for a simulation
%starting from a $W_0=5$ King model with $N=8192$, $10~\%$ primordial
%binaries and $m_{BH}=0.014M$ (solid line), compared to a similar
%simulation without the BH (dotted line). The snapshot is taken at time
%$t=22~t_{rh}(0)$. \label{fig:binW5}}
%\end{figure}

%\begin{figure}
%\resizebox{\hsize}{!}{\includegraphics{f12.eps}} \caption{Spatial distribution of the binaries for the
%two simulations plotted in Fig.~\ref{fig:binW5}, with snapshots taken
%at $t=11~t_{rh}(0)$ and at $t=22~t_{rh}(0)$. \label{fig:binW5spatial}}
%\end{figure}

%%%%%%%%%%%%%%%%%%%%%%%%%%%%%%%

%%%%%%%%%%%%%%%%%%%%%%%%%%%%%%%%%%%%%%%%%%%%%%%%%%%%

\section{Escapers}

The N-body systems in our simulations, despite being isolated, lose
mass, i.e. some particles acquire a positive energy and are
consequently removed from the system (if they are unbound and reach a
distance of $20~r_h$ from the center of system). There are two main
contributions to this process. One is diffusion of particles in the
energy space, that leads to the evaporation of particles with binding
energy very close to $0$ (that translates into a small residual
velocity at infinity, which we call ``ejection velocity''). The other
is due to strong interactions of binaries with other stars and/or with
the IMBH with the consequent ejection of a high velocity particle (the
slingshot effect, see \citealt{aar05}).

\begin{figure}
\resizebox{\hsize}{!}{\includegraphics{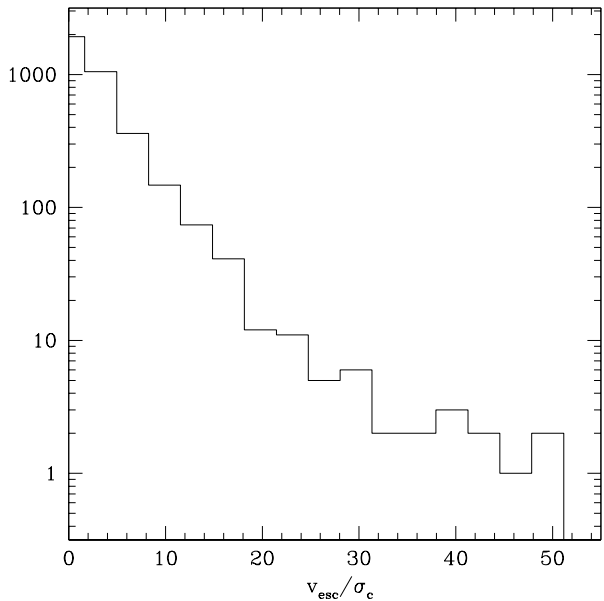}}
\caption{Distribution of the escape velocity
  (in units of the core velocity dispersion at the escape time) as
  measured in a sample of $5$ simulations with $N=8192$, $10\%$
  primordial binaries, and $m_{BH}=0.014$. If applied to a typical
  star cluster (where $\sigma_c \approx 10km/s$), this plot suggests
  that the combined action of an IMBH plus a population of primordial
  binaries can lead to the ejection of stars at several hundreds
  $km/s$.\label{fig:esc}}
\end{figure}

In our Plummer run with $N=8192$, $f=10\%$ with $m_{BH}=0.014 M$ we
find that a fraction of $3 \cdot 10^{-3}$ of the stars of the cluster
leave the system in a relaxation time and that $\approx 1/3$ of them
have ejection velocities that exceed by at least five times the core
velocity dispersion. Among these $54$ were binaries; in addition one
triple was ejected from the system. A distribution of the ejection
velocities, as measured in a series of 5 simulations with $N=8192$, is
shown in Fig.~\ref{fig:esc}.

These results, especially for the high velocity tail of the
distribution, need to be confirmed by more realistic simulations that
include a mass spectrum and account for possible stellar collisions
during close encounters, but they suggest that the interaction of IMBH
with a significant population of binaries may efficiently lead to the
production of a number of ultra fast stars, i.e stars with an ejection
velocity of several hundreds $km/s$.

{\add The ejection velocity resulting from a binary-IMBH interaction
is limited to be of the order of the orbital velocity of the binary
members with respect to their center of mass (e.g., see
\citealt{yu03}). This implies that for main sequence stars the maximum
ejection velocity is of the order of the escape velocity from the star
(of order of a few hundreds $km/s$) and this is approximately
independent on the mass of the stars due to the (quasi) linear
mass-radius relationship along the main sequence. Therefore only binaries
with at least one highly compact member (a white dwarf, a neutron star
or a stellar mass BH) can lead to the production of hypervelocity
stars (i.e. stars with velocity exceeding $1000 km/s$) in consequence
of tidal shredding due to an IMBH. From the observational point of view
this decreases the likelihood of a globular clusters origin for main
sequence hypervelocity stars. The expected leading production channel
is in this case an interaction with a supermassive black hole to be
produced, as initially proposed by \citet{hil88} and recently
investigated by \citet{yu03}, \citet{gin06} and
\citet{bau06}. Unfortunately the possibility of identifying
observationally ejected stars with velocities below $1000$ km/s does
not appear very promising. In fact the several hundreds of km/s range
of velocities is shared also by stars in the disk of our galaxy, so an
accurate measure of the proper motion would be required to trace back
to the center of a globular cluster the trajectory of a candidate main
sequence ejected star.}

%%%%%%%%%%%%%%%%%%%%%%%%%%%%%%%%%%%%%%%%%%%%%%%%%%%%%%%%%%%%%%%%%%%%%%%%%%%
%%%%%%%%%%%%%%%%%%%%%%%%%%%%%%%%%%%%%%%%%%%%%%%%%%%%%%%%%%%%%%%%%%%%%%%%%%%

\section{Discussion and conclusions}\label{sec:conclusion}

In this paper we have presented for the first time the results of
direct N-body simulations of the evolution of a star cluster with
a population of primordial binaries that harbors a central IMBH, with
a mass of the order of $1~\%$ of the total mass of the system. In
order to begin to analyze the basic evolutionary processes we have
restricted our simulations to isolated systems of stars of equal mass,
neglecting at this stage stellar evolution and physical collisions.

The environment around the BH turns out to be of great interest from a
dynamical point of view. It represents a laboratory where complex
interactions between hierarchical systems take place. Around the BH we
observe usually one or more stars tightly bound to it. With a
significant population of primordial binaries, it frequently happens
that a binary approaches the center of the system.  Interactions between
these subsystems are important not only from a dynamical point of view,
but also because they form a factory to produce exotic objects, such as
tight (X-ray emitting) binaries and high velocities escapers, which we
observe in high numbers in our simulations. 

From a computational point of view the complex interactions between
binaries and the BH have proved to be very challenging for current
state of the art N-body algorithms. In our simulations we had to
resort to several restarts of the computation to fine tune the
integration parameters in order to achieve energy errors conservation
below $5 \cdot 10^{-3}$ at the end of the simulation and below $5
\cdot 10^{-4}$ per dynamical time. Clearly, new regularization methods
have to be developed to provide both more accuracy and higher
stability. This will allow to perform automatic runs, a necessary
condition to employ a more systematic study of systems with IMBH and
binaries.

{\add We have shown that a BH with a mass of around $1~\%$ of the cluster
mass can produce most of the energy required to fuel the expansion of
the star cluster, so that the evolution of the large scale structure
of a star cluster with primordial binaries and a central BH is
remarkably similar to that attained in the absence of binaries (see
\citealt{bau04a}). Even so, the evolution of the population of
primordial binaries is strongly influenced by the BH:} binaries are
disrupted primarily within the influence sphere of the BH, but we note
a significant lack of binaries tidally disrupted by the BH with
respect to predictions based on the loss cone theory. These
theoretical estimates only take into account interactions between the
binaries and the BH starting from the density of binaries in the
core. Instead we observe that most of the binaries are destroyed by
three and four body encounters with other stars once they enter the
influence sphere of the BH and before they have the chance to venture
close enough to the BH to be tidally shredded.
%{\add This effect may have important consequences on the
%estimate of the rate of gravitational and X-Ray events associated with
%tidal disruptions of binaries.}

These three and four body encounters within the influence sphere of
the BH often lead to the ejection of stars with proper motions easily
exceeding 10 times the core velocity dispersion of the cluster. We
observe events with velocities up to 50 times the core velocity
dispersion, that is of the order of $10 km/s$ for a typical globular
clusters. Binaries within the influence sphere of an IMBH can therefore
originate a significant population of high velocity stars ejected from
the system, {\add whose presence could be detected observationally. Tidal
shredding of binaries with main sequence components limits the
ejection velocity to a few hundreds $km/s$, but higher velocities can
be obtained if at least one member of the binary is a compact object
(white dwarf, neutron star, or stellar mass BH). One related
intriguing, although highly speculative possibility is indeed the
dynamical origin of a $\approx 1000 km/s$ pulsar candidate from the
globular cluster remnant IRS13, which is considered to host an IMBH
\citep{wan06}.}

{\add In principle, an indirect evidence for the presence of an IMBH in
 an old globular cluster could be obtained by observing the
 distribution of binaries: observational detection of an unexpectedly
 large fraction of only moderately hard binaries in the core could be
 used {\add as circumstantial evidence} for the presence of an IMBH in
 old star clusters, i.e. in those systems whose age would imply a
 depletion in the central region of binaries with binding energy below
 $\approx 15~kT$.  With our runs we have shown that these binaries are
 able to survive in significant numbers when an IMBH is present. For a
 typical globular cluster this would imply to be able to measure in
 the core the ratio of binaries with semi-major axis of about $a
 \gtrsim 0.5~AU$ to those with $a \approx 0.05~AU$, that appears to be
 particularly sensitive to the presence of an IMBH (see
 Fig.~\ref{fig2}). This relative measure has the advantage of being
 relatively independent from the precise value of the primordial
 binary ratio (clearly unknown for observed cluster), but it is
 extremely challenging for the present observational techniques of
 binary detection. This would in fact mean to be able not only to
 detect a significant number of binaries in the dense core environment
 of a globular cluster, but also to accurately measure the semi-major
 axis of these binaries down to a fraction of AU. In addition, as we
 have mentioned in Sec. 4, if softer primordial binaries are
 preferentially born in the center of the cluster, this diagnostic
 fails.

A more promising observational evidence for the presence of an IMBH in
old, relaxed, globular clusters can instead be given by a simpler
measure, i.e. the ratio of the core to half mass radius. This quantity
is set by the efficiency of energy generation in the core of the
cluster and is robust with respect to changes in the details of the
initial conditions. Systems made of single stars only have a very
small $r_c/r_h$ ratio ($r_c/r_h \approx 0.01$), as no efficient energy
production mechanism is available. When primordial binaries are
present, $r_c/r_h$ is up to almost one order of magnitude larger and
the ratio depends only on the initial fraction of binaries, saturating
at $f \gtrsim 10\%$ (see \hth and \thh). Accounting for a small
N-dependence of the ratio, a typical old globular cluster with
primordial binaries is expected to have $r_c/r_h \lesssim 0.05$ (see
also \citealt{fre06}). Most of the observed globular clusters have
instead much larger values for $r_c/r_h$, located in the range
$[0.1;1]$ \citep{fre06}. If we were to assume that old globular
clusters host an IMBH, then we would naturally obtain $r_c/r_h \approx
0.3$ which is in good agreement with the observations.}

Clearly, a quantitative comparison with observations would require us
to proceed beyond the simple models presented here. To start with, we
would need to include a realistic number of stars to avoid possible
biases introduced by the scaling of the ratio of the BH mass to that
of single stars. In addition, the introduction of a realistic initial
mass function is likely to modify the concentration of binaries in the
center of the system due to mass segregation, and stellar evolution will
also influence the distribution of binary binding energy.  However,
the main effects presented in this paper are likely to be present, at
least in qualitative ways, in more detailed realistic simulations.

\section{Acknowledgments}

We are indebted to Sverre Aarseth for providing his code NBODY-6 with
ad-hoc modifications in order to make possible running the simulations
presented in this work. We are grateful to the referee for
constructive suggestions that have improved the paper. We thank Holger
Baumgardt and Jun Makino for interesting discussions. This work is
supported in part by the Grants-in-Aid of the Ministry of Education,
Culture, Sports, Science and Technology, Japan, (14079205; MT, EA, SM)
and by a Grant-in-Aid for the 21st Century COE {\lq\lq}Center for
Diversity and Universality in Physics" (SM). P.H. thanks
Prof. Ninomiya for his kind hospitality at the Yukawa Institute at
Kyoto University, through the Grants-in-Aid for Scientific Research on
Priority Areas, number 763, "Dynamics of Strings and Fields", from the
Ministry of Education, Culture, Sports, Science and Technology,
Japan. The numerical simulations have been performed on the Condor
cluster at the Institute for Advanced Study.

%%%%%%%%%%%%%%%%

\appendix
\section{Numerical Accuracy}\label{sec:error}

The Aarseth's NBODY6 code represent one of the few state of the art
integrators for the direct integration of collisional N-body systems.
Its performance and accuracy have been extensively tested (see
\citealt{aar03}). The code employs a series of automatic internal
accuracy tests at the level of individual timesteps and forces,
treatment of hierarchical systems and, of course, conservation of the
total integral of motions. If one of these tests fails, the code
resumes the computation from a previously saved snapshot of the system
with improved integration parameters, to try to obtain an accuracy
within the level required. If this is not possible, the code
eventually stops the computation, returning an error message.

\begin{figure}
\resizebox{260pt}{!}{\includegraphics{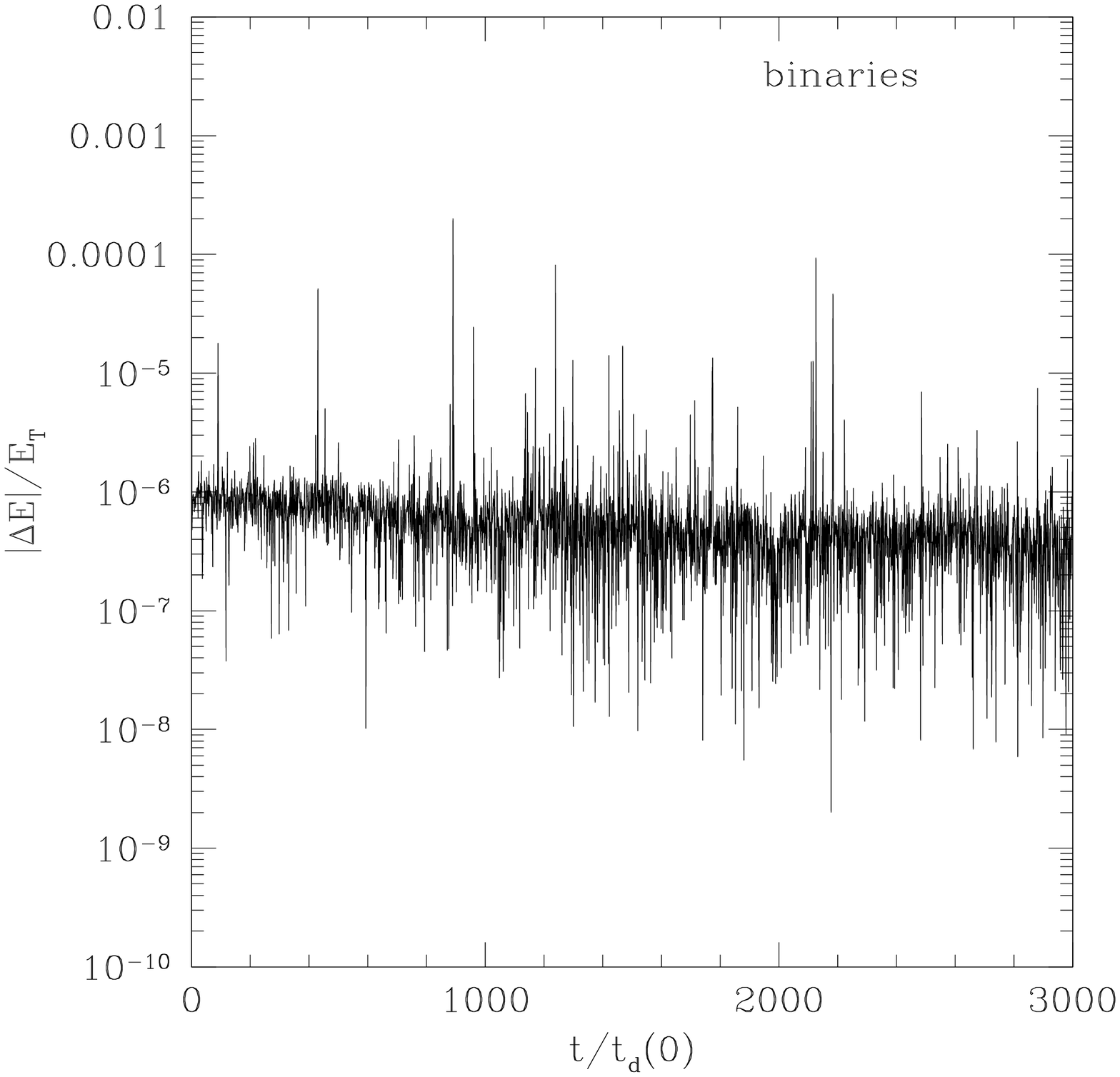}\includegraphics{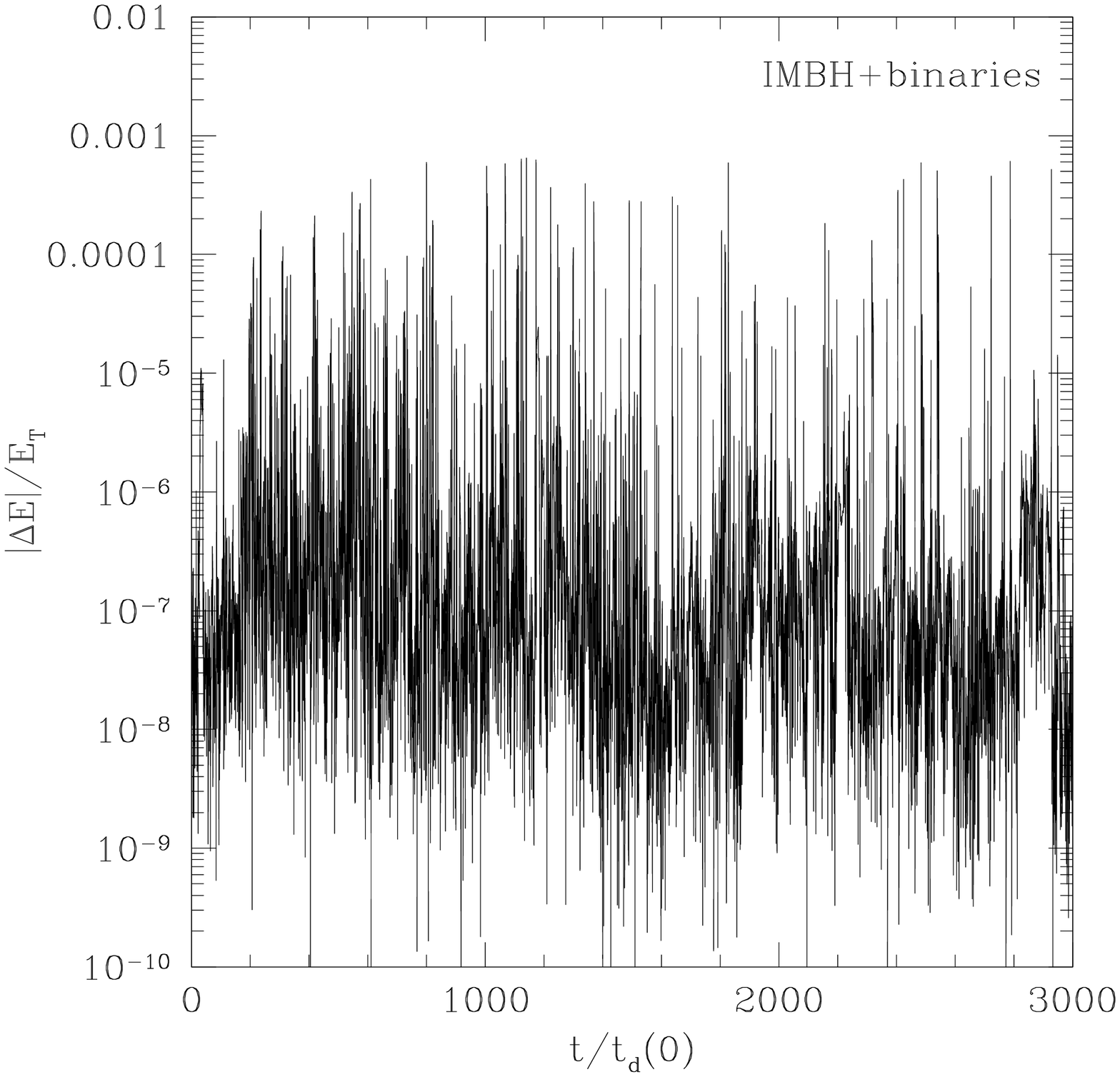}} \caption{Relative total energy
  conservation per dynamical time $t_d$ for $N=8192$ in a run with $10
  \%$ primordial binaries (left panel) and in a run with $10\%$
  primordial binaries and an IMBH with mass $1.4 \%$ of the total mass
  of the system (right panel). The maximum allowed error per dynamical
  time is set at $5 \cdot 10^{-4}$; all our runs are stopped if the
  total energy error increases above $5 \cdot 10^{-3}$.\label{fig:error}}
\end{figure}

Typically, for a system with $N \approx 10^4$, the code can guarantee
relative errors in the conservation of the total energy at the level
of $10^{-8}$ per dynamical time. The introduction of a population of
primordial binaries increases significantly the dynamical range of the
computation, as the shortest dynamical timescale to be resolved may be
of the order of days, compared a dynamical time of a few million of
years for a typical globular cluster. This is reflected with a
relative energy error of the order of $10^{-6}$ per dynamical time
(see Fig.~\ref{fig:error}), that still guarantees an excellent
conservation of the energy at the end of the simulation (the errors
tend to have random signs so that the total error increases
approximately with the square root of the time). The code has a
similar performance also for runs that include single stars plus a
IMBH (see \citealt{bau04a,bau04b}).

The introduction of an IMBH in simulations with primordial binaries
introduces additional challenges for the code, as in this case several
multiple interacting systems have to be treated in the sphere of
influence of the IMBH. Despite the improvements introduced to increase
the performance of the code (see Sec.\ref{sec:ns}), significantly
greater errors with respect to the standard runs for the code happened
during our simulations (see Fig.~\ref{fig:error}), leading to a number
of restart of the run performed manually in order to fine tune the
integration options, depending on the kind of error message received.

To ensure reliable results we set a maximum allowed energy error per
dynamical time of $5 \cdot 10^{-4}$ and we definitely stop a
simulation that reaches a total energy error of $5 \cdot
10^{-3}$. This threshold value is typically reached after about 25
initial half mass relaxation times (thus at a time longer than the
Hubble time for a typical globular cluster).

%%%%%%%%%%%%%%%%%%%%%%%%%%%%%%%%%%%%%%%%%%%%%%%%%%%%%%%%%%%%%%%%%%%%%%%%%%%%%%%5
\bsp

\label{lastpage}


\begin{thebibliography}{}

\bibitem[Aarseth(2003)]{aar03}
Aarseth S.,  2003, {Gravitational N-body Simulations}.~ Cambridge
University Press

\bibitem[Aarseth(2005)]{aar05} Aarseth, S. 2005, To appear in Few-Body Problem, Annales Universitatis Turkuensis, C. Flynn, ed astro-ph/0511565

\bibitem[Albrow et al.(2001)]{alb01} Albrow, M. D. et al. 2001, \apj,
		       559, 1060

\bibitem[Bahcall \& Wolf(1976)]{bah76}
Bahcall, J. N. and Wolf, R. A. 1976, \apj, 209, 214

\bibitem[Baumgardt et al.(2003a)]{bau03a}
Baumgardt, H., Hut, P., Makino, J., McMillan, S. and Portegies Zwart S.
2003, \apj, 582, L21

\bibitem[Baumgardt et al.(2003b)]{bau03b}
Baumgardt, H., Makino, J., Hut, P., McMillan, S. and Portegies Zwart,
S. 2003, \apj, 589, 25

\bibitem[Baumgardt et al.(2004a)]{bau04a} Baumgardt, H., Makino, J., and
		       Ebisuzaki T. 2004, \apj, 613, 1133

\bibitem[Baumgardt et al.(2004b)]{bau04b} Baumgardt, H., Makino, J., and
		       Ebisuzaki T. 2004, \apj, 613, 1143

\bibitem[Baumgardt et al.(2006)]{bau06} Baumgardt, H., Gualandris,
A. and Portegies Zwart, S. 2006, \mnras, 372, 174

\bibitem[Baumgardt et al.(2005)]{bau04c} Baumgardt, H., Makino, J., and Hut, P. 2005, \apj, 620, 238

\bibitem[Bellazzini et al.(2002)]{bel02} Bellazzini, M.,
		       Pecci, F.~P., and Messineo M. 2002, \apj, 123, 1509

\bibitem[Casertano \& Hut(1985)]{cas85} Casertano, S. and Hut, P. 1985, \apj, 298, 80

\bibitem[Ferrarese \& Merritt(2000)]{fm00} Ferrarese, L. and Merritt, D. 2000, \apj, 539, L9

\bibitem[Fregeau et~al.(2003)]{fre03}
{Fregeau} J.~M.,  {G{\" u}rkan} M.~A.,  {Joshi} K.~J.,    {Rasio} F.~A.,  2003,
  \apj, 593, 772

\bibitem[Fregeau \& Rasio(2006)]{fre06} {Fregeau} J.~M. and {Rasio} F.~A 2006, \apj, submitted, astro-ph/0608261

\bibitem[\protect\citeauthoryear{{Gao}, {Goodman}, {Cohn} \& {Murphy}}{{Gao}
  et~al.}{1991}]{gao91}
{Gao} B.,  {Goodman} J.,  {Cohn} H.,    {Murphy} B.,  1991, \apj, 370, 567

\bibitem[Gebhardt et al.(2000)]{geb00}
{{Gebhardt}, K. and {Bender}, R. and {Bower}, G. and {Dressler}, A. and 
	{Faber}, S.~M. and {Filippenko}, A.~V. and {Green}, R. and {Grillmair}, C. and 
	{Ho}, L.~C. and {Kormendy}, J. and {Lauer}, T.~R. and {Magorrian}, J. and 
	{Pinkney}, J. and {Richstone}, D. and {Tremaine}, S.} 2000, \apj, 539, L13 

\bibitem[Gebhardt et al.(2002)]{geb02}
Gebhardt, K., Rich, R.~M., and Ho, L.~C. 2002, \apj, 578, L41

\bibitem[Gebhardt et al.(2005)]{geb05}
Gebhardt, K., Rich, R.~M., and Ho, L.~C. 2005, \apj, 634, 1093

\bibitem[Gerssen et al.(2003)]{ger03}
Gerssen, J., van der Marel, R. P., Gebhardt, K., Guhathakurta, P.,
Peterson R. C., and Pryor, C., 2003, \aj, 125, 376

\bibitem[Giersz \& Heggie(1994)]{gie94}
{Giersz} M.,  {Heggie} D.~C.  1994, \mnras, 268, 257

\bibitem[Giersz \& Spurzem(2000)]{gie00}
{Giersz} M.,  {Spurzem} R.  2000, \mnras, 317, 581

\bibitem[Ginsburg \& Loeb(2006)]{gin06} Ginsburg, I. and Loeb, A. 2006, \mnras, 368, 221

\bibitem[\protect\citeauthoryear{{Heggie} \& {Mathieu}}{{Heggie} \&
  {Mathieu}}{1986}]{hm86}
{Heggie} D.~C.,  {Mathieu} R.~D.,  1986, in LNP 267: The Use of
  Supercomputers in Stellar Dynamics {Standardised Units and Time Scales}.
p 233

\bibitem[Heggie \& Aarseth(1992)]{heg92}
{Heggie} D.~C.,  {Aarseth} S.~J.,  1992, \mnras, 257, 513

\bibitem[\protect\citeauthoryear{{Heggie} \& {Hut}}{{Heggie} \&
  {Hut}}{2003}]{heg03}
{Heggie} D.,  {Hut} P.,  2003, {The Gravitational Million-Body Problem: A
  Multidisciplinary Approach to Star Cluster Dynamics}.~ Cambridge
University Press

\bibitem[Heggie, Trenti \& Hut(2006)]{heg05} Heggie, D.C., Trenti, M., and
  Hut, P. 2006 \mnras, 368, 677

\bibitem[H{\' e}non(1965)]{hen65} H{\' e}non, M. 1965, Ann. Astr., 28, 62

\bibitem[Hills(1988)]{hil88} Hills, J.G. 1988, Nat, 331, 687 

\bibitem[Hurley et al.(2005)]{hur05} Hurley, J.R., Pols, O.R.,
  Aarseth, S.J. and Tout, C.A. 2005, \mnras, 363, 293

\bibitem[Hut(1989)]{hut89} Hut, P., 1989,
  Fundamental Timescales in Star Cluster Evolution, in  {\it Dynamics of
  Dense Stellar Systems.} ed. David Merritt (Cambridge University
  Press), pp. 229-236.

\bibitem[\protect\citeauthoryear{{Hut}, {McMillan}, {Goodman}, {Mateo},
  {Phinney}, {Pryor}, {Richer}, {Verbunt} \& {Weinberg}}{{Hut}
  et~al.}{1992}]{hut92}
{Hut} P.,  {McMillan} S.,  {Goodman} J.,  {Mateo} M.,  {Phinney} E.~S.,
  {Pryor} C.,  {Richer} H.~B.,  {Verbunt} F.,    {Weinberg} M.,  1992, \pasp,
  104, 981

\bibitem[Kustaanheimo \& Stiefel(1965)]{ks64} Kustaanheimo, P. and Stiefel, E. 1965, J. Reine Angew. Math., 218, 204 

%\bibitem[Makino \& Aarseth(1992)]{mak92} {Makino}, J. and {Aarseth}, S.~J. 1992, \pasj, 44, 141

\bibitem[McMillan \& Hut(1994)]{mcm94}
{McMillan} S.,  {Hut} P.,  1994, \apj, 427, 793

\bibitem[McMillan et~al.(1990)]{mcm90}
{McMillan}, S.,  {Hut}, P.,    {Makino}, J.  1990, \apj, 362, 522

%\bibitem[Mikkola \& Aarseth(1996)]{mik96} {{Mikkola}, S. and {Aarseth}, S.~J.} 1996, Celestial Mechanics and Dynamical Astronomy, 64, 197

%\bibitem[Mikkola \& Aarseth(1998)]{mik98} {{Mikkola}, S. and {Aarseth}, S.~J.} 1998, New. Astron., 3, 309

\bibitem[Peebles(1972)]{pee72} Peebles, P.~J. 1972, \apj, 178, 371


\bibitem[Portegies Zwart et al.(2004)]{por04} Portegies Zwart, S.~F., Baumgardt, H., Hut, P., Makino, J., and McMillan, S.~L.~W. 2004, \nat, 438, 724


\bibitem[Portegies Zwart \& McMillan(2004)]{por04b}
{Portegies Zwart} S.~F.,  {McMillan} S.~L.~W.,  2004, Proc. of
conference on Massive Stars in Interacting Binaries, Canada,
N. St-Louis and A. Moffat eds. (astro-ph/0411188)


\bibitem[Pfahl(2005)]{pfa05} Pfahl, E. 2005, \apj, 626, 849


\bibitem[Spitzer(1987)]{spi87} Spitzer, L.  1987, Dynamical Evolution of
    Globular Clusters, Princenton University Press, Princenton

\bibitem[Tremaine et al.(2002)]{tre02} {{Tremaine}, S. and {Gebhardt}, K. and {Bender}, R. and {Bower}, G. and 
	{Dressler}, A. and {Faber}, S.~M. and {Filippenko}, A.~V. and 
	{Green}, R. and {Grillmair}, C. and {Ho}, L.~C. and {Kormendy}, J. and 
	{Lauer}, T.~R. and {Magorrian}, J. and {Pinkney}, J. and {Richstone}, D.
	} 2002, \apj, 574, 740

\bibitem[Trenti, Heggie \& Hut(2006)]{tre06} Trenti, M., Heggie, D.C., and
  Hut, P. 2006, \mnras, submitted (astro-ph/0602409)

\bibitem[Vesperini \& Chernoff(1994)]{ves94} Vesperini, E. \& Chernoff, D.~F.\ 1994, \apj, 431, 231 

\bibitem[Yu \& Tremaine(2003)]{yu03} Yu, Q. and Tremaine S. 2003, \apj, 599, 1129

\bibitem[Wang et al.(2006)]{wan06} Wang, Q.~D.,  {Lu}, F.~J. \&
  {Gotthelf}, E.~V. 2006, \mnras, 367, 937

\bibitem[Zhao \& Bailyn (2005)]{zha05} Zhao, B. \& Bailyn, C.~D. 2005, \aj, 129, 1934 

\end{thebibliography}
\end{document}